\documentclass[epj,twocolumn]{webofc}
\usepackage[varg]{txfonts}   
\usepackage{units}
\woctitle{International Symposium on Very High Energy Cosmic Ray Interactions 2014}
\begin{document}
\title{Charm production in \textsc{SIBYLL}}
%
%

\author{F. Riehn\inst{1}\fnsep\thanks{\email{felix.riehn@kit.edu}}
  \and R. Engel \inst{1} \and
  A. Fedynitch\inst{1,2} \and
  T. K. Gaisser\inst{3} \and
  T. Stanev\inst{3}}

\institute{Karlsruher Institut f\"{u}r Technologie, Institut f\"{u}r
  Kernphysik, Postfach 3640, 76021 Karlsruhe, Germany \and CERN
  EN-STI-EET, CH-1211 Geneva 23, Switzerland \and Bartol Research
  Institute, Department of Physics and Astronomy, University of
  Delaware, Newark, DE 19716, USA }

\abstract{\textsc{SIBYLL}~2.1 is an event generator for hadron
  interactions at the highest energies. It is commonly used to analyze
  and interpret extensive air shower measurements. In light of the
  first detection of PeV neutrinos by the IceCube collaboration the
  inclusive fluxes of muons and neutrinos in the atmosphere have
  become very important. Predicting these fluxes requires
  understanding of the hadronic production of charmed particles since
  these contribute significantly to the fluxes at high energy through
  their prompt decay. We will present an updated version of
  \textsc{SIBYLL} that has been tuned to describe LHC data and
  extended to include the production of charmed hadrons.  }
\maketitle
\section{Introduction}
\label{sec-1}
\textsc{SIBYLL}~\cite{Fletcher:1994bd} is a hadronic interaction model
that is widely used in air shower simulations. It is available as one
of the standard hadronic interaction models for high energy in the
simulation packages AIRES, CORSIKA, CONEX and SENECA. \textsc{SIBYLL}
is also used for calculating atmospheric lepton
fluxes~\cite{Fedynitch:2012fs}.

The current version of the model is
\textsc{SIBYLL}~2.1~\cite{Ahn:2009wx}. It is designed to allow simulation
of hadronic interactions in the energy range from $\unit[\sqrt{s} \approx 10]{GeV}$
up to $\unit[400]{TeV}$. At the time of tuning the parameters of this model,
TeVatron data ($\sqrt{s}\sim\unit[2]{TeV}$) were the
highest energy measurements available.
In this work we present a new version of \textsc{SIBYLL} tuned to
accelerator data including those from LHC. In addition, this version
has been extended to include a phenomenological model of the production of 
charmed hardons.

In the first section we describe the updates of the model motivated
by LHC data. This includes the refit of the cross section parameters,
the extension of the fragmentation model to increase
baryon pair production, and the update of the parton distribution
functions. In the second section we describe the model of charm
production, how the parameters are adjusted to describe data and what
conclusions can be drawn from applying the model in calculations of
the inclusive flux of atmospheric leptons.

\section{LHC updates}
\label{sec-2}

Before discussing the update of the model it is worthwhile to mention
that \textsc{SIBYLL}~2.1 already describes the general characteristics of
hadronic interactions at $\unit[7]{TeV}$ remarkably well (see dashed
blue histogram in Fig.~\ref{fig:etad_chd} or the review by d'Enterria
et al.~\cite{dEnterria:2011kw}).

\subsection{Proton proton cross section}
\label{sec-21}
The hadron-proton cross section in \textsc{SIBYLL} follows from unitarizing
hard and soft cross section contributions, separated by an energy dependent cutoff in
$p_\perp$, and terms due to diffraction dissociation. More details on the structure
of the model can be found in the publication for
version~2.1~\cite{Ahn:2009wx}.

\begin{figure}
  \centering
  \includegraphics[width=\columnwidth,clip]{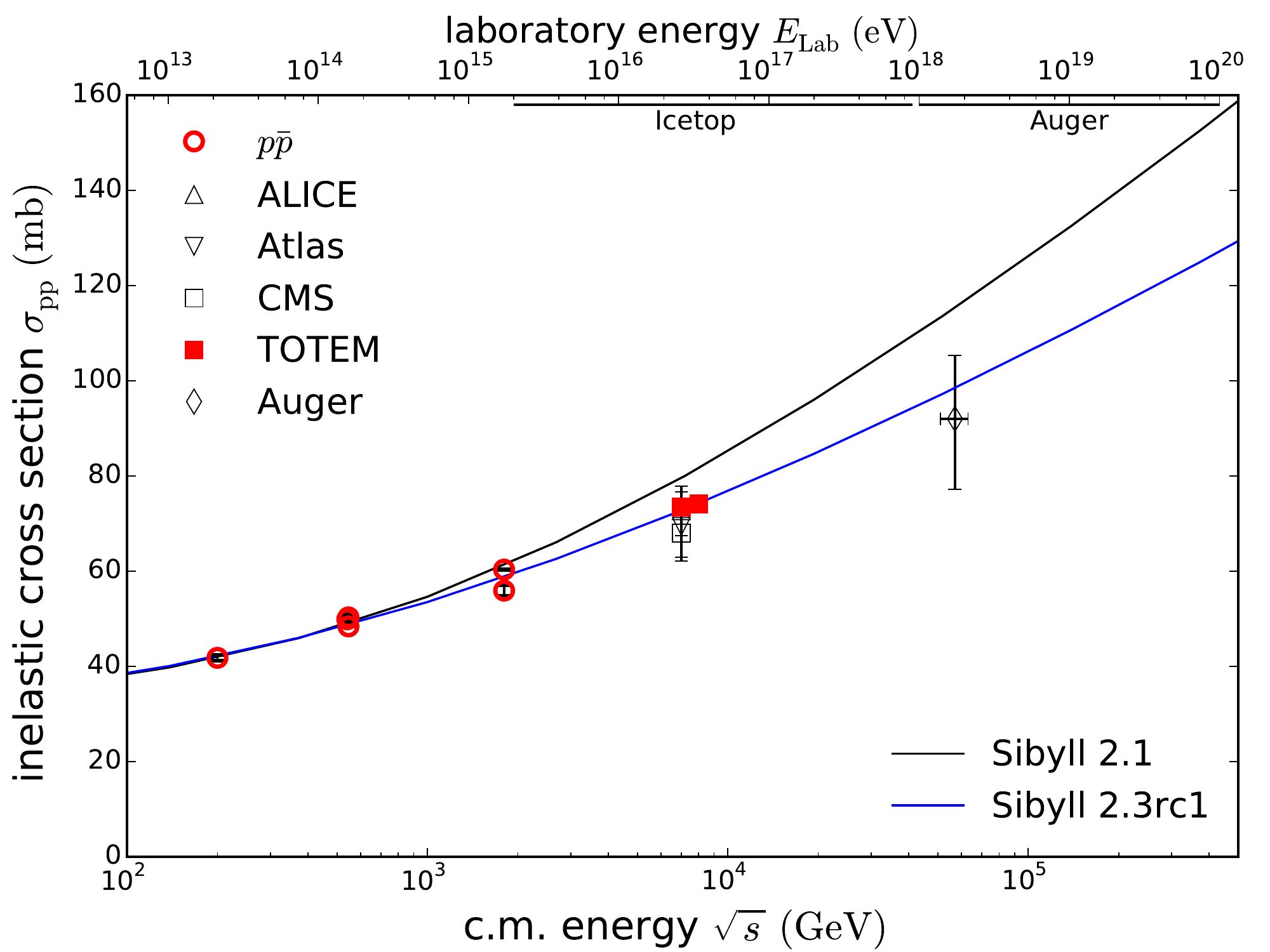}
  \caption{Inelastic $p$-$p$ cross section in \textsc{SIBYLL}. The
    updated cross section is shown in blue, the old version is in
    black. The red squares are the measurements by
    TOTEM~\cite{Antchev:2011vs}. The black diamond at the highest
    energy is the estimate from the Auger
    Observatory~\cite{Auger:2012wt,Abraham:2004dt}. The second energy
    axis shows the equivalent laboratory energy for $p$-$p$
    interactions as applicable to air shower detectors (one proton at
    rest). The measurement ranges of the IceTop air shower
    array~\cite{IceCube:2012nn} and the Pierre Auger
    Observatory~\cite{Abraham:2004dt} are indicated by black lines.  }
  \label{fig:xsctn}       
\end{figure}

Measurements at LHC suggest (see Fig.~\ref{fig:xsctn}) that
\textsc{SIBYLL}~2.1 overestimates the cross section at high
energies. The inelastic cross section measured in the TOTEM
experiment, which has the highest precision for a measurement of the
total cross section at LHC, is
$\unit[73.5^{+1.9}_{-1.4}]{mb}$~\cite{Antchev:2011vs} whereas
\textsc{SIBYLL} predicts $\unit[80]{mb}$. The rise of the $pp$ cross
section beyond $\unit[1]{TeV}$ is mainly driven by hard parton scattering (hard
minijets).

In \textsc{SIBYLL} an eikonal approximation is used to combine the
parametrization of soft scatterings with the perturbative calculation
of the minijets into an unitary amplitude, which then defines the
total and elastic cross sections. The size of the soft and hard
contributions in this formalism depends on the size of the particular
cross section and the profile function.

In order to make the inelastic cross section compatible with the TOTEM
result without changing the hard cross section (calculated within
QCD), the profile function of the distribution of the hard partons
in transverse (impact parameter) space
has been made more narrow so that
peripheral collisions are less likely to produce minijets.

The downside of this approach is that central collisions now exhibit
very high densities of interacting partons (profile functions are normalized),
which means that some events will have a large number of minijets and
consequently a large number of final state particles produced. This
effect will produce a tail in the multiplicity distribution that is
not observed in data. However central collisions are rare so the
average multiplicity and most other observables are still compatible
with the measurements~\cite{dEnterria:2011kw}.

Since our goal is a model capable of describing interactions a decade
and more higher in center-of-mass energy, the effects of high parton densities have
to be considered, even if the mean multiplicity still agrees with
current experiments. A microscopic model of parton density
\textit{saturation} could limit the number of scatterings in central
collisions and thereby repair the multiplicity. In the current model,
saturation is implemented only in an impact parameter independent way
as an energy-dependent lower $p_{\perp}$-cutoff for the minijets.

In addition to changing the hard profile function we adjust the
parameters of the soft cross section parametrization to fit the
$p$-$p$ and $\bar{p}$-$p$ cross section data. 

Since the proton profile also enters the meson-nucleon cross sections,
we refit the parametrization of the soft contribution there as well.

The resulting cross section is shown in Fig.~\ref{fig:xsctn} as a blue
line. The old cross section for comparison is shown as a black solid
line. The data point of the highest energy is the estimation of the
$p$-$p$ cross section from air shower measurements at the Auger
Observatory at energies of about $\unit[\sqrt{s} =
  57]{TeV}$~\cite{Auger:2012wt,Abraham:2004dt}. This value has not
been used to fit the cross section in \textsc{SIBYLL} and therefore
can be seen as an indication that the extrapolation by the model is
reasonable.

\subsection{Baryon production}
\label{sec-22}

\begin{figure}
  \centering
  \includegraphics[width=\columnwidth,clip]{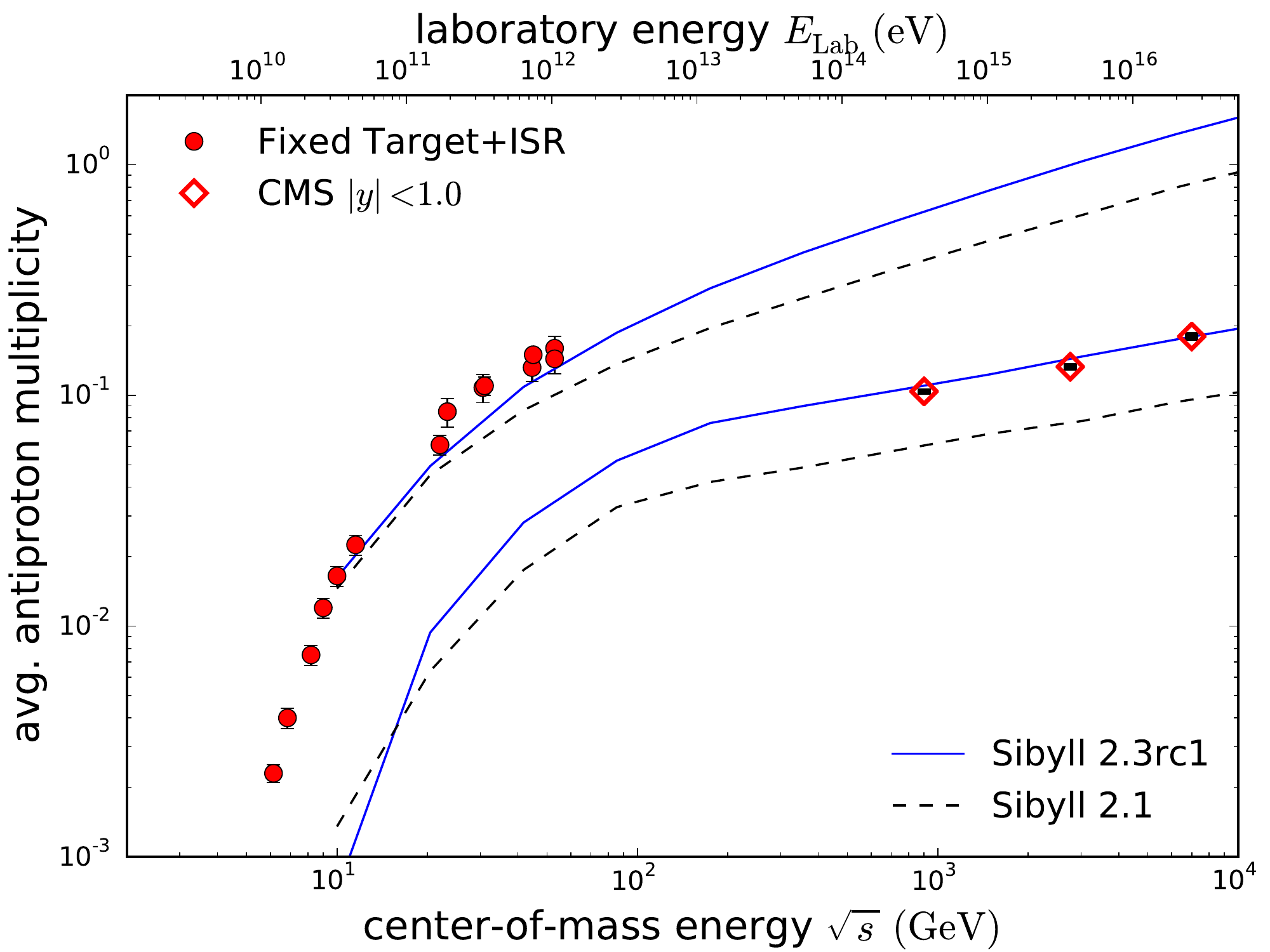}
  \caption{Average antiproton multiplicity as a function of
    center-of-mass energy. The low energy data are a compilation of
    fixed target and ISR experiments that cover the full phase space
    or were extrapolated to full phase space~\cite{Antinucci73}. The CMS
    data~\cite{Chatrchyan:2012qb} are taken in a phase space region with
    $|y|<1.0$. PHENIX~\cite{Adare:2011vy} data are taken in the range
    $|\eta|<0.35$. The prediction by the models are shown for the full
    and CMS phase spaces only. \textsc{SIBYLL}~2.1 is shown as
    dashed line, the updated version as solid line.}
  \label{fig:pbarmult}
\end{figure}

Particle production in interaction models primarily depends on the
implementation of the fragmentation process. Fragmentation is a
non-perturbative process so the rates of particle production cannot be
calculated from first principles, which means the parameters in the
model have to be set by comparison with experiment.

In \textsc{SIBYLL} \textit{string
  fragmentation}~\cite{Sjostrand:1987xj} is used as the fragmentation
model. The string model simplifies hadronization by assuming a uniform
energy density in the color field stretched between two partons which
eventually is split in two by quark-antiquark pair production. The
splitting is continued until the remaining energy is just enough to
form two hadrons. Baryons are produced by introducing
diquark-antidiquark pairs instead of $q\bar{q}$ pairs. The probability
of producing a diquark pair rather than a quark pair ($P_{\rm{q/qq}}$)
in a string breakup is the parameter that controls baryon pair
production. In version~2.1 it is set to $\unit[0.04]{}$.

For simplicity, only two string classes are distinguished in
\textsc{SIBYLL}: the 2 string configuration for the $2 \rightarrow 2$
sea parton scattering and two single strings connecting valence
quarks/diquarks. The essential difference between the two is that the
latter configuration has valence flavor attached to the string ends, where as
the former is in total flavor neutral. This distinction is necessary
to describe the differences between the forward/backward regions and
the central region of phase space.

The result of this treatment of baryon production in
\textsc{SIBYLL}~2.1 for the antiproton multiplicity is shown in
Fig.~\ref{fig:pbarmult} as dashed black lines together with a
compilation of data. The multiplicity for full phase space, typically
measured in fixed target experiments at low energies, is shown in the
upper set of lines whereas the multiplicity in the central region ($|
\eta | < \unit[2]{}$), the region typically accessible in collider
experiments, e.g.\ CMS~\cite{Chatrchyan:2012qb}, is shown in the lower
set. The current model describes the threshold at low energies well
but is not capable of describing the central, high energy data at the
same time.

In order to allow for a meaningful extrapolation to high energies,
instead of introducing an arbitrary energy dependent parametrization
for $P_{\rm{q/qq}}$, one can relate the baryon production frequency
to soft and semihard interactions,
whose energy dependence is different, to increase the baryon
production mainly at high energy. With the minijet cross section
being derived from perturbative QCD the extrapolation to higher energy
is then given by the model and, at low energy,
threshold effects due to the large mass of the
baryon pairs are important.

Furthermore minijets mostly produce particles in the central region
which is exactly where the high energy data by CMS reveal a deficit
for \textsc{SIBYLL}~2.1. This assumption is supported by the
observation of the ratio of antiprotons to charged pions compared to
the central charged multiplicity (see e.g.\ Fig.~15 in
Ref.~\cite{Chatrchyan:2012qb}).

The simplest possible coupling of the diquark production parameter
to minijets is to choose a different but fixed value of $P_{\rm{q/diq}}$
in the fragmentation of minijets in comparison to all other fragmentation
processes. The resulting model describes the data much better (solid blue line in
Fig.~\ref{fig:pbarmult}), especially in the central region.

Measurements of baryon production at LHC energies that cover the
forward phase space could test the assumptions made in this model.

\subsection{Transverse momentum of minijets}

In \textsc{SIBYLL}~2.1 the momentum fractions that determine the
kinematics of the minijets are taken from an effective parton density
function~\cite{Combridge:1983jn}
\begin{equation}
f(x) ~=~ g(x) \,+\, \frac{4}{9} \left[ q(x) \,+\, \bar{q} (x) \right] ~,~
\label{eqn:eff_pdf}
\end{equation}
where $g(x)$ and $q(x)$ are parametrized according to Eichten et al.
(EHLQ)~\cite{Eichten85a} and the quark distribution function includes
contributions from three light flavors ($u$, $d$, and $s$) and the valence quarks.

\begin{figure}
  \includegraphics[width=\columnwidth]{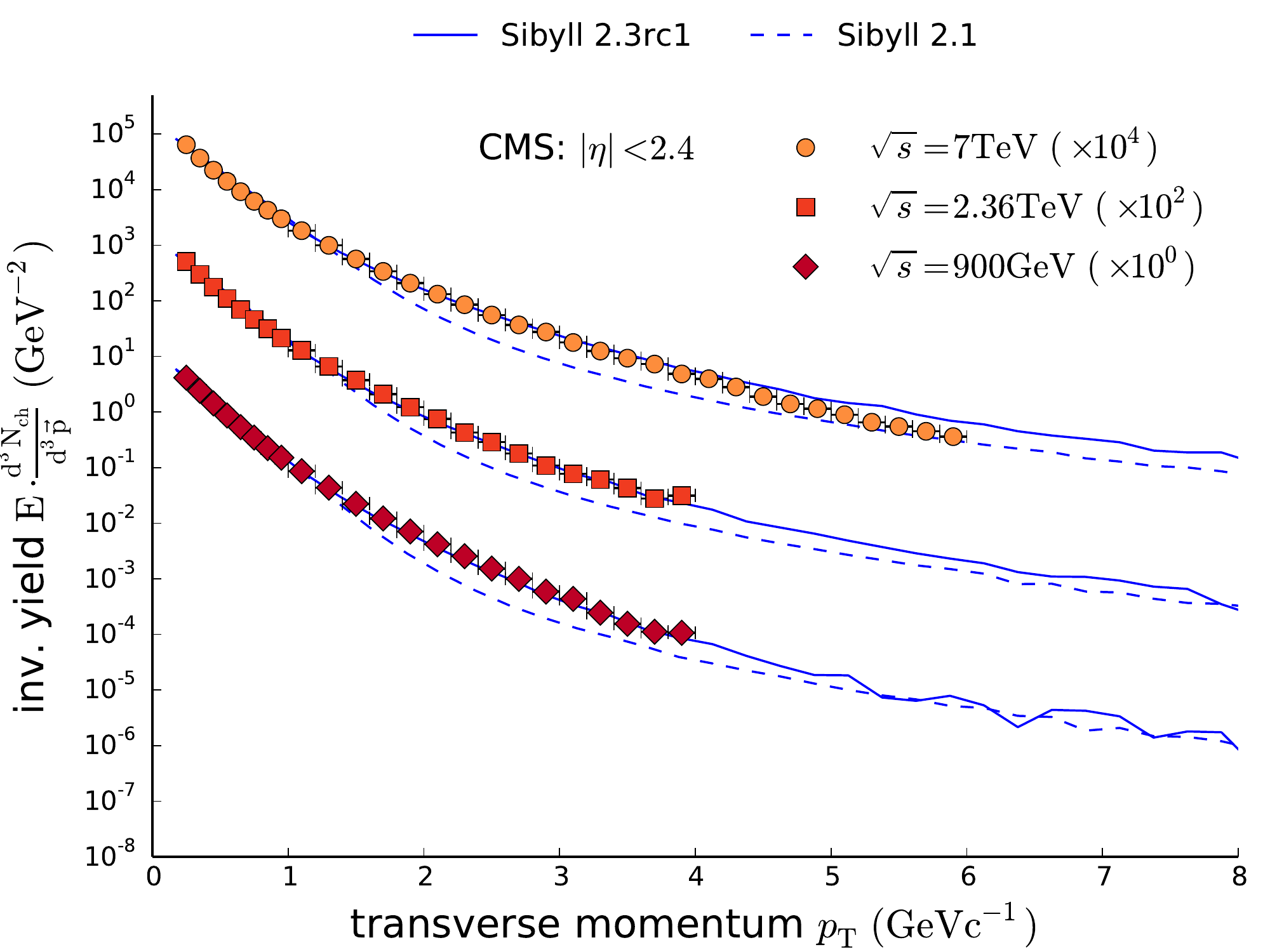}
  \caption{Inclusive cross section for charged particles as function
    of the transverse momentum. The results obtained with the old and
    new versions of \textsc{SIBYLL} are compared with CMS data at
    different
    c.m.\ energies~\cite{Khachatryan:2010xs,Khachatryan:2010us}.
  \label{fig:ptd_chd}}
\end{figure}

In the updated version the same effective parton distribution function (PDF)
is used but the quark and gluon contributions are sampled from the
same PDF parametrizations (GRV~\cite{Gluck:1994uf,Gluck98a})
that are used in the calculation of the hard minijet cross section.

The main difference between these parametrizations is the behavior at
low $x$ which, in the case of the GRV parametrization, is much
steeper.

In combination with the correction of a mistake in the definition of
the $p_{\perp}^{\rm{min}}$ the steeper PDFs give a better description
of the spectra in the range of intermediate transverse momenta
($\unit[2]{}-\unit[5]{GeV/c}$) than in \textsc{SIBYLL}~2.1, see
Fig.~\ref{fig:ptd_chd}.

\subsection{Other updates}
\label{sec-23}

Other general and more technical aspects of the model that have been
updated but are not discussed here are: the transverse momentum
acquired in the soft scattering of valence quarks as well as in the string
fragmentation is now sampled from an exponential transverse mass
distribution rather than a Gaussian as in the previous version.

Another aspect of direct importance to air shower predictions
is the enhanced forward production of vector mesons with respect
to pseudoscalar mesons (pions) in meson nucleon
interactions~\cite{Agababyan:1990df}. Since this mechanism has a
large influence on muon production in air showers it has been implemented
in the new version of \textsc{SIBYLL}.

Furthermore the implemented Glauber model for the 
calculation of the different cross sections
(total, elastic, diffractive, and quasi-elastic) in
hadron-nucleus interactions has been extended to include a consistent
treatment intermediate low-mass excitations, leading to enhanced screening
effects~\cite{Glauber:1955qq}.

\subsection{Comparison to data}
\label{sec-24}

To show the compatibility of the updated model with experimental data
we look at the charged particle \textit{pseudorapidity
  distribution}. The advantage of this observable is that it is very
sensitive to the details of the parton level interaction structure and
kinematics ($n_{\rm jets},x_{i}$) as well as to the subsequent
fragmentation process (${\rm d}N^{\rm ch}_{\rm string}/{\rm d}\eta$).

\begin{figure}
  \centering
  \includegraphics[width=\columnwidth,clip]{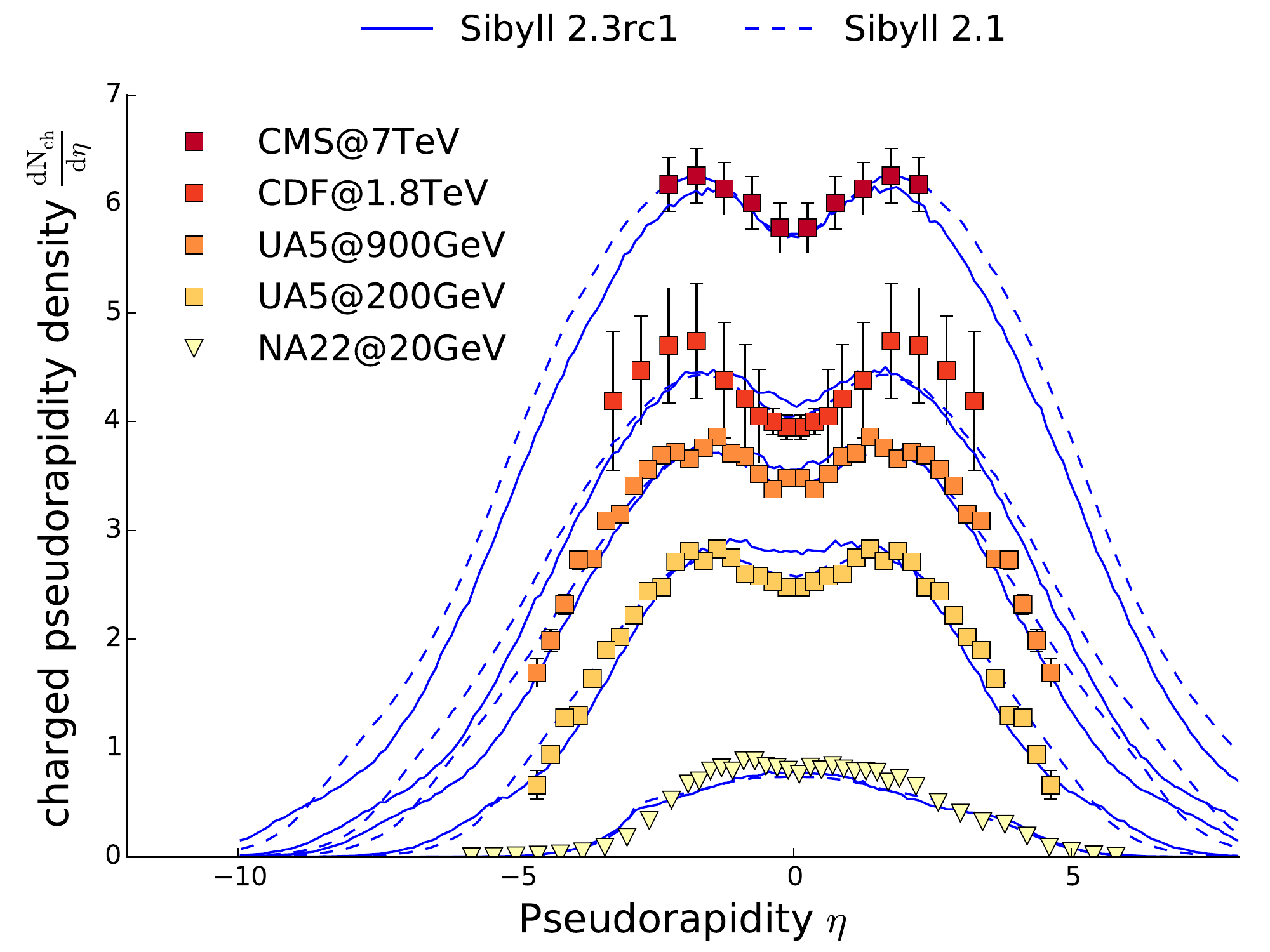}
  \caption{Pseudorapidity distribution of charged particles. The data
    are from
    NA22~\cite{Adamus:1988xc},UA5~\cite{Alner:1986xu},CDF~\cite{Abe:1989td}
    and CMS~\cite{Khachatryan:2010us}. The prediction by
    \textsc{SIBYLL}~2.1 is shown by the dashed line, the one for the
    updated model by the solid line.}
  \label{fig:etad_chd}
\end{figure}

The changes introduced in the cross section are expected to increase
the central multiplicity at energies beyond $\unit[1]{TeV}$ whereas
the increased baryon production, due to the higher mass of baryonic
particles, can be expected to lead to an overall decrease in the
multiplicity. Fig.~\ref{fig:etad_chd} shows that both effects
approximately cancel one another at LHC energy and that also the updated
model (solid blue histogram) describes the CMS data well.

\section{Charm quark extension}
\label{sec-3}

\textsc{SIBYLL}~2.1 is limited to the production of particles
containing $u$, $d$, and $s$ quarks. In version 2.2c of \textsc{SIBYLL} it was
shown that a simple phenomenological extension of the fragmentation model, based on the
family connection between strange and charmed hadrons, can account for
the production of charmed particles at low
energy~\cite{Ahn:2011wt}. In this approach the normalization is set by
the rate at which charm quarks appear relative to strange quarks
$P_{c}=\unit[0.004]{}$.

\subsection{New charm model}
\label{sec-31}
Due to the high mass of the charm quark the production of charmed
hadrons in the fragmentation process is suppressed by a large factor.
Instead the dominant channel is the direct production of charm quarks
in parton-parton scattering. In the context of QCD the leading
contribution $g g \rightarrow c\bar c$ is often referred to
as \textit{QCD gluon-gluon fusion}~\cite{Lourenco:2006vw}. The momentum transfer of
the reaction due to the charm quark mass $Q^2>Q_{\rm{min}}\sim 2m_{c}$
means that the process can be expected to be calculable within perturbation theory.

The \textsc{SIBYLL} event generator includes only the dominant terms of
hard parton-parton scattering at high energy and does not distinguish between the
hadronization of the different parton configurations. All
parton-parton scattering processes are fragmented into hadrons through
an unflavored two string configuration, similar to 2 scattered gluons
(usually referred to as hard minijets).

To account for the dominating hard scattering contribution the charm
quark fraction is increased in the fragmentation of the hard
minijets. To keep the threshold behavior at low energy the charm
quark fraction is suppressed exponentially in low mass
strings. Specifically
\begin{equation}
  P^{i}_{c\bar{c}} = P^{i}_{c,0} \exp{ \left( -
    \frac{m_{\textrm{eff}}}{\hat{s}} \right ) } ~,~
  \label{eq:chm_par}
\end{equation}
where $\hat{s}$ is the invariant mass of the scattering partons and
$m_{\rm eff}=\unit[20]{GeV^2}$ is the effective mass scale. To account for
string configurations of higher order charm production is not limited to
the end of the strings but extends over the whole string. This part
of the phenomenological model for charm production 
is referred to as perturbative component.

Next to the dominant contribution from hard scattering, experiments
have shown that there is an asymmetry in charm production in the
fragmentation region (i.e.\ at large
$x_{\rm F}$)~\cite{Alves:1996rz,Aitala:1996hf}, which suggests a
contribution from charm production in soft interactions. Two models,
which can be used to explain
this forward production of heavy flavor, are the \textit{intrinsic
charm} model~\cite{Brodsky:1980pb} and the \textit{flavor excitation}
model~\cite{Combridge:1978kx}.

In \textsc{SIBYLL} we chose a model which could represent either
mechanism by adding charm quarks to any string attached to soft
scattered partons as well (non-perturbative component).  These will
include valence strings which, due to the large momentum fraction and
the attached flavor of the valence quarks, are able to produce the
observed asymmetry at large $x_{\rm F}$.

\subsection{Tuning the charm parameters}

The values of the parameters in Eq.~\ref{eq:chm_par} are adjusted
separately for the perturbative and non-perturbative contribution.
The perturbative part is tuned to describe the $p_{\perp}$-spectra of
$D$ mesons measured by the ALICE~\cite{ALICE:2011aa} and
LHCb~\cite{Aaij:2013mga} experiments in central phase space, since this
is where its contribution is expected to be dominant
(Fig.~\ref{fig:hg_chm}). The parameters for the soft contribution are
set to account for the missing production at low energies
(Fig.~\ref{fig:lw_chm}).

\begin{figure*}
  \centering
  \includegraphics[width=0.8\textwidth,clip]{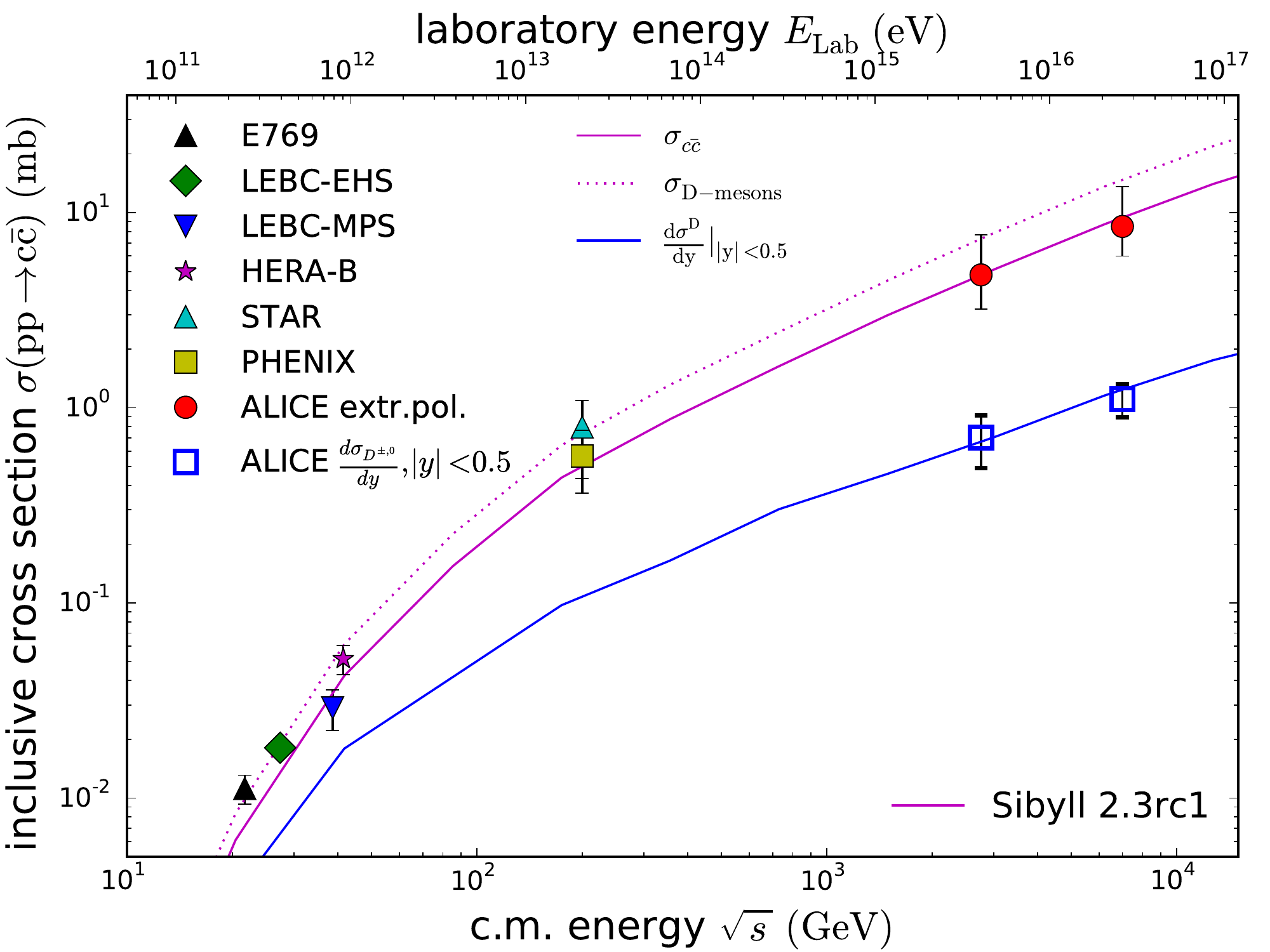}
  \caption{Inclusive charm and $D$-meson cross sections as a function of
    c.m.\ energy. The data at low energy are $D$-meson cross sections in
    fixed target
    experiments~\cite{Alves:1996rz,AguilarBenitez:1988sb,Ammar:1988ta,Zoccoli:2005yn}. The
    measurements at the highest energies are $c\bar{c}$ from
    ALICE~\cite{Abelev:2012vra,ALICE:2011aa}. Here data are shown
    extrapolated to full phase space (red circles) and visible only
    (blue empty squares). At intermediate energies the data taken at
    RHIC by the STAR~\cite{Adamczyk:2012af} and
    PHENIX~\cite{Adare:2010de} experiments are shown (also
    extrapolated). The model prediction for the inclusive $c\bar{c}$ cross
    section is shown by the solid line, the prediction for the
    production of $D$-mesons is shown by the dotted line.}
  \label{fig:charm_tot}       
\end{figure*}

\begin{figure}
  \includegraphics[width=\columnwidth,clip]{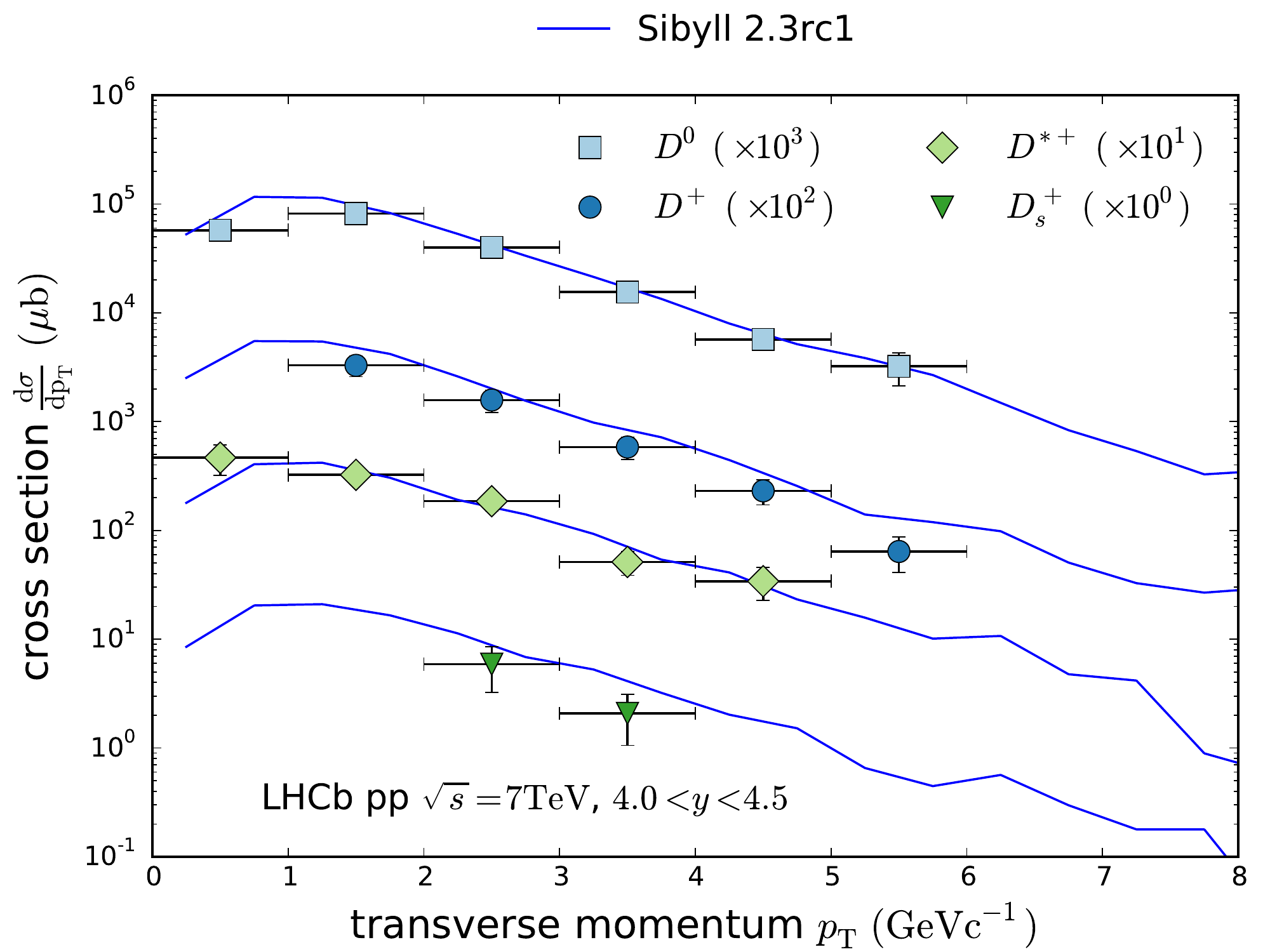}
  \caption{Transverse momentum spectrum of different types of $D$-mesons
    in the rapidity interval $4.0<y<4.5$. Data were taken at
    $\sqrt{s}=\unit[7]{TeV}$ with the LHCb
    detector~\cite{Aaij:2013mga}. \label{fig:hg_chm}}
\end{figure}

\begin{figure}
  \includegraphics[width=\columnwidth,clip]{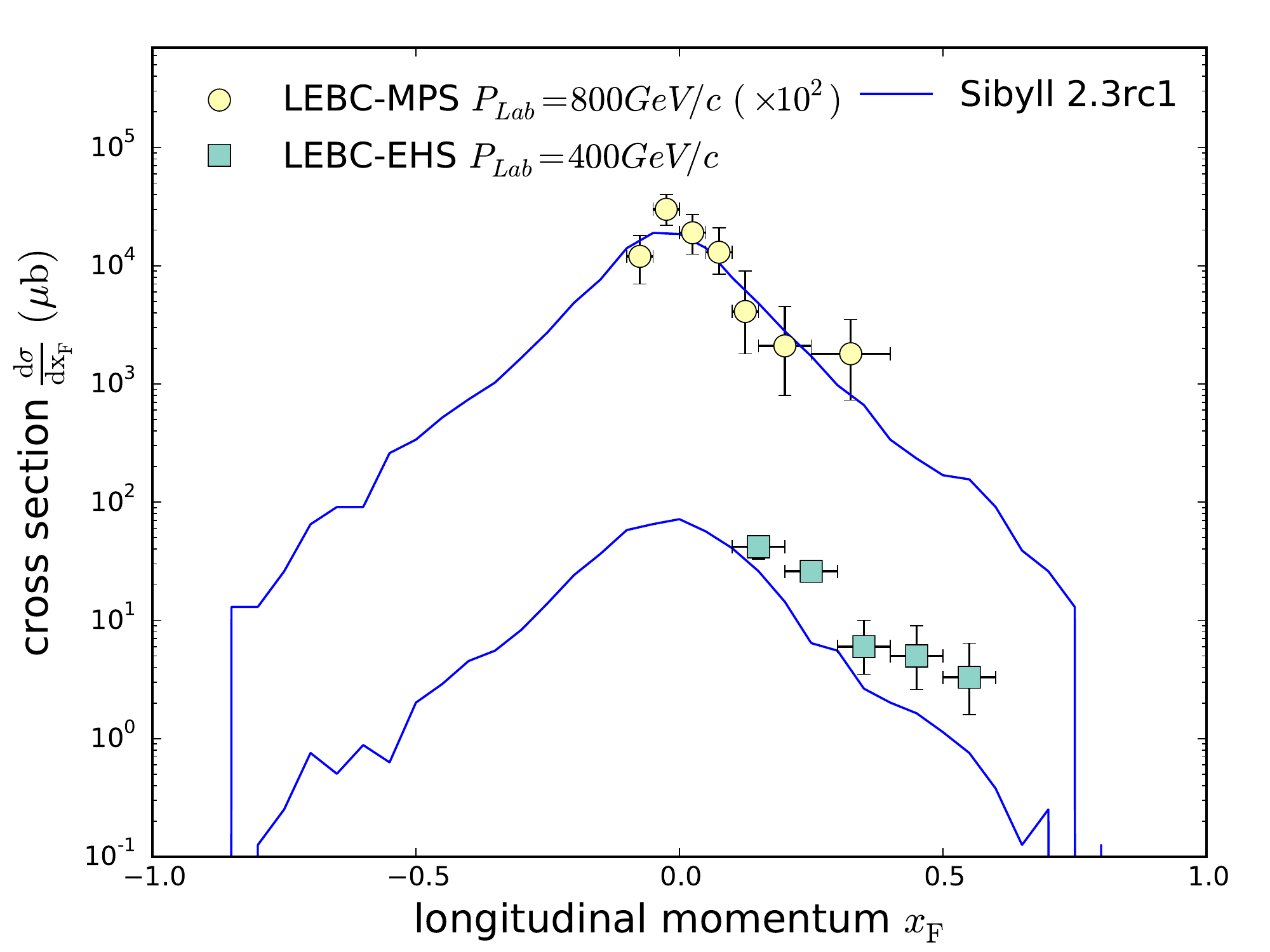}
  \caption{Feynman-$x$ spectra of charged $D$ mesons in $p$-$p$ fixed
    target interactions with
    $P_{\rm{Lab}}=\unit[400]{GeV}$~\cite{AguilarBenitez:1988sb} and
    $\unit[800]{GeV}$~\cite{Ammar:1988ta}.\label{fig:lw_chm} }
\end{figure}

In Fig.~\ref{fig:charm_tot} the cross section for inclusive charm
production is shown as a function of the center-of-mass energy. The
ALICE data include an extrapolation from central to total phase space.
The cross section  for $D$ meson production that is measured directly by
ALICE is shown by the lower blue points and lines. The dotted line
represents the inclusive $D$ meson production cross section without
subtracting the decays of resonances of higher mass, e.g. $D^{*}$. It
is shown here because the low energy measurements are not corrected
for this either.

The resulting model correctly describes the rise of the inclusive
charm cross section with energy and reproduces the spectra at
different energies.

\subsection{Discussion}

\begin{figure}
  \includegraphics[width=\columnwidth,clip]{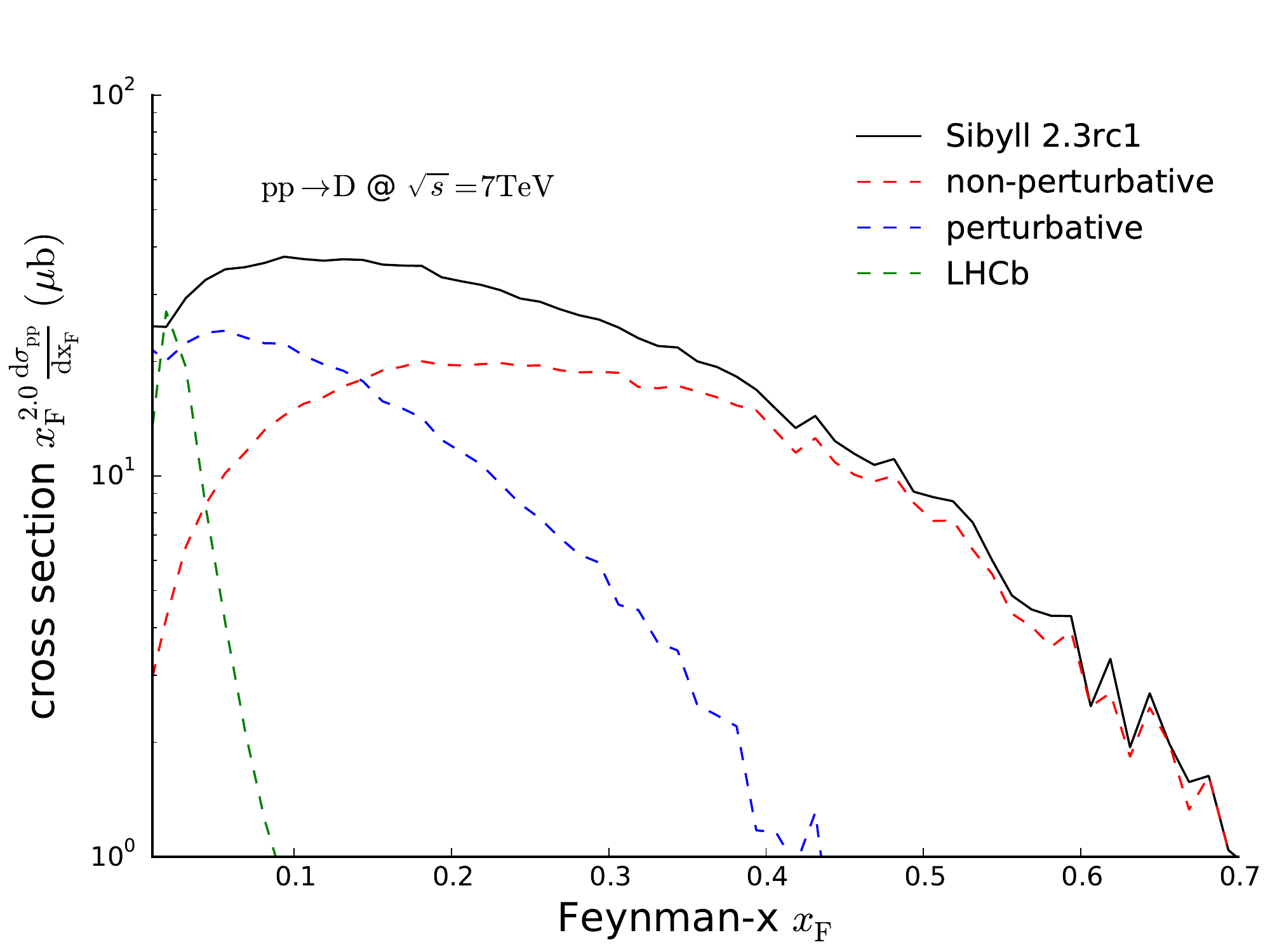}
  \caption{Weighted spectrum for $D$-mesons in \textsc{SIBYLL} at
    $\sqrt{s}=\unit[7]{TeV}$. The contributions from the perturbative
    and non-perturbative model components are shown by the blue and red lines,
    respectively. Note the negligible contribution to the energy
    spectrum from the phase space covered by the LHCb experiment
    ($2.5<y<4.5$, green line). \label{fig:ps}}
\end{figure}

\begin{figure}
  \includegraphics[width=\columnwidth,clip]{{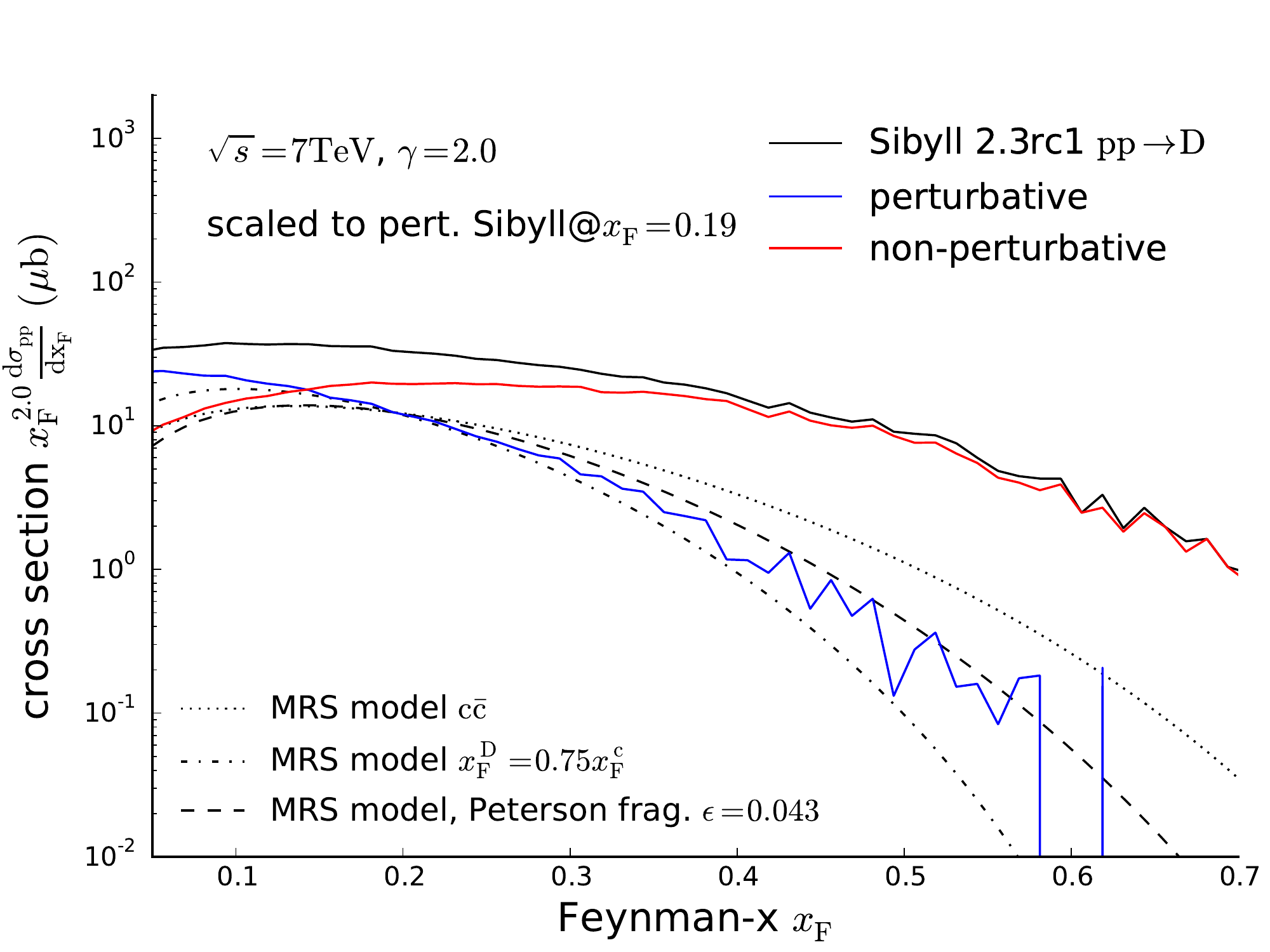}}
  \caption{Comparison of the weighted spectrum of $D$-mesons in
    \textsc{SIBYLL} and the MRS model~\cite{Martin:2003us}. The MRS
    spectra are shown for different fragmentation
    assumptions.\label{fig:mrs}}
\end{figure}

Charmed particles were introduced to the model because they are
expected to contribute significantly to the inclusive flux of
atmospheric muons and neutrinos at high
energy~\cite{Gaisser:2002jj}. The inclusive fluxes can be obtained by
solving the corresponding cascade equation~\cite{Gaisser:1990vg}.  The
results depend on the spectrum weighted moments
\begin{equation}
  Z_{pD} = \int_{0}^{1} x_{\rm L}^{\gamma}\,\frac{{\rm d}n}{{\rm d}x_{\rm{L}}}
  \,{\rm d}x_{\rm{L}} ~,~ \label{eqn:zfact}
\end{equation}
where $x_{\rm L}$ is the energy fraction of the considered final state
particle in the laboratory frame,
$\frac{{\rm d}n}{{\rm d}x_{\rm{L}}}$ is the hadronic production spectrum,
and $\gamma$ is the power law index of the integral all-nucleon spectrum of 
cosmic rays. With $1.7<\gamma<2.3$ it is evident from Eq.~\ref{eqn:zfact}
that the contribution to the lepton flux is largest for charm production at
large $x_{\rm L}$.

Unfortunately, particle production at large $x_{\rm L}$ is difficult
to study at high energy. So far there are no experiments that cover
this part of phase space and are capable of particle identification
(PID) at the same time. The most forward detector at the LHC with PID
capabilities is LHCb ($2.5<y<4.5$). For \textsc{SIBYLL} the
contribution from this phase space to $Z_{pD}$ is only about $\unit[10]{\%}$
at $\sqrt{s}=\unit[7]{TeV}$ (integrating the green line in
Fig.~\ref{fig:ps}). The prediction of the contribution of charmed
particles to the inclusive neutrino and lepton fluxes therefore are
not well constrained by the measurements at the LHC. In addition,
large-$x_{\rm L}$ production is dominated by soft, non-perturbative
processes, so the prediction can not be well constrained by
theoretical arguments either.

In Fig.~\ref{fig:mrs} a comparison of the weighted energy
spectrum, i.e.\ the integrand in Eq.~\ref{eqn:zfact}, for $D$ mesons in
$p$-$p$ interactions between the model by Martin, Ryskin and Stasto
(MRS)~\cite{Martin:2003us} and \textsc{SIBYLL} is shown. The MRS model
is a perturbative calculation of the charm production that is extended
to low momentum fractions using additional assumptions and accounting for
saturation. The energy of the comparison is 
$\unit[7]{TeV}$ and the index, with which the spectrum is weighted, is
$\gamma=2$. To compare the shapes of the distributions the models are
scaled such that MRS and the perturbative
component in \textsc{SIBYLL} are equal at $x_{\rm{F}}=\unit[0.19]{}$.
One can see that the MRS model and the perturbative component in
\textsc{SIBYLL} show a similar behavior. The main difference between the
models is that \textsc{SIBYLL} predicts additional charm production
from the non-perturbative component that is dominating in the forward
direction.

A detailed calculation of the atmospheric lepton fluxes using the
model discussed here can be found in our second
contribution~\cite{Fedynitch:2014af} to this conference. In that paper,
the role of the all-nucleon spectrum and the atmosphere are
discussed as well.

It should be mentioned that the entire discussion here was focused on
proton-proton interactions. What matters for the atmospheric fluxes
is the charm production in nucleon-nucleus interactions. In principle, the
model is implemented such that central (perturbative) charm production
should scale approximately with the number of binary interactions,
while forward charm production scales with the number of projectile
participants. In practice these scaling expectations are not really satisfied
because of additional energy-momentum constraints. Given the strong
dependence of the atmospheric fluxes on the forward production nuclear
screening effects in this region could have a large effect. For
central production, the measurements confirm the binary
scaling~\cite{fortheALICE:2013ica}.

\section{Summary and Outlook}

An improved version (2.3rc1) of the hadronic interaction model
\textsc{SIBYLL} has been presented. The current status of the 
update of the $p$-$p$ cross section,
the extension of the fragmentation model to describe
increased baryon production, and the new charm production model have
been described in more detail. The perturbative component of the
charm model was found to be compatible with the analytic MRS
calculation. It was also shown that the experimental data currently
available on charm production do not directly restrict the predictions
for the inclusive muon and neutrino fluxes. Only indirectly, by comparing
model predictions with charm measurements in phase space regions
covered also at colliders, constraints on atmospheric lepton fluxes
can be derived.

In the future we plan to estimate how large the uncertainty
in the atmospheric fluxes due to the limited phase space coverage of
the measurements is. This can be achieved by looking for a set of
parameters in the charm model that either minimizes or maximizes
the forward charm yield
while still being compatible with experimental data.

\section*{Acknowledgments}
It is a pleasure to acknowledge many inspiring and fruitful discussions
with colleagues of the IceCube, KASCADE-Grande, and Pierre Auger Collaborations.
This work is supported in part by the Helmholtz Alliance for Astroparticle
Physics HAP, which is funded by the Initiative and Networking Fund of
the Helmholtz Association.



%
%
%

%
%

\end{document}